\documentclass[%
 aip,
 amsmath,amssymb,
 reprint,%
]{revtex4-1}

\usepackage{graphicx}
\usepackage{dcolumn}
\usepackage{bm}

\usepackage[utf8]{inputenc}
\usepackage[T1]{fontenc}
\usepackage{mathptmx}
\usepackage{etoolbox}
\usepackage[version=4]{mhchem} 
\usepackage{float}
\usepackage[section]{placeins}
\makeatletter
\def\@email#1#2{%
 \endgroup
 \patchcmd{\titleblock@produce}
  {\frontmatter@RRAPformat}
  {\frontmatter@RRAPformat{\produce@RRAP{*#1\href{mailto:#2}{#2}}}\frontmatter@RRAPformat}
  {}{}
}%
\makeatother
\begin{document}

\preprint{AIP/123-QED}

\title{Post Annealing Crystallization behavior of RF Sputtered Yttrium Iron Garnet thin films on Si/SiO$_2$ patterned substrates}
\author{M. Roman}
\affiliation{Electrical \& Computer Engineering, The University of New Mexico, Albuquerque, NM, USA}
 \affiliation{ 
Center for High Tech Technology Materials (CHTM), Albuquerque, NM, USA
}%
\author{A. Siddiqui}
\affiliation{ 
Sandia National Laboratories, Albuquerque, NM, USA
}%
\author{T. Busani}
\affiliation{Electrical \& Computer Engineering, The University of New Mexico, Albuquerque, NM, USA}
 \affiliation{ 
Center for High Tech Technology Materials (CHTM), Albuquerque, NM, USA
}%

\date{\today}

\begin{abstract}
Yttrium Iron Garnet YIG (Y$_3$Fe$_5$O$_{12}$), is a commonly used material for magnonic devices due to its crystal and chemical structure, which makes the material highly ferromagnetic and enables long-range magnon propagation. Magnonic devices were fabricated by depositing a 390 nm thick thin film of YIG, using low vacuum RF sputtering, on Si substrates with a 240 nm buffer layer of SiO$_2$. Two sets of devices were used to study the effect of the Si/SiO$_2$ interface on the YIG. The first set features patterned hole pairs on the Si/SiO$_2$, which was created using fluorine etching. Patterned samples were used as seed nucleation points to study the crystallization behavior. The second set was a non-patterned Si/SiO$_2$ with YIG deposited uniformly on the top. Post-deposition recrystallization of the YIG film was accomplished in a horizontal furnace under O$_2$ atmosphere, between 750 °C and 850 °C. EDS shows that YIG is off-stoichiometric and rich in iron and oxygen, as deposited. There was also no interaction between the Si/SiO$_2$ substrate and the YIG, as there was no change in post annealing stoichiometry. XRD patterns show that full crystallization occurred at the 800 °C for 1 h annealing for the patterned samples, while for the non-patterned sample, crystallization continues to progress at 750 °C after 3 h. The thin film YIG appears to form an intra-phase at temperatures below 800° C, and segregates into Fe$_2$O$_3$ and Y$_2$O$_3$. Raman spectroscopy showed that YIG started to crystallize at the seed nucleation points and propagated to the entire thin film YIG. Finally, we noticed that longer annealing times or high-temperature annealing present strong tensile stress on the YIG, which also results in the formation of polycrystalline rather than single-crystal YIG. Ellipsometry measurements are consistent with this conclusion; the real part of the refractive index of the amorphous sample was n$\approx$ 2.6, while for the crystallized sample annealed at 800 °C for 1 h, it was n$\approx$ 5.1. This work demonstrates a pathway for designing suspended magnonic devices. The study highlights the importance and advantages of crystallizing patterned YIG films. By patterning devices with a SiO$_2$ buffer layer, depositing YIG via RF sputtering, and subsequently crystallizing the films in a furnace, we establish a fabrication route toward devices that can be suspended. Although further optimization of stoichiometry is required, achieving precise compositional control would enable the realization of fully suspended and released YIG devices that can be transferred onto alternative substrates.
\end{abstract}
\maketitle
\section{Introduction}
Magnonics is a relatively new and emerging area of study that relies on magnons or spin waves as information carriers \cite{ref1}. Magnonics have many advantages over traditional electronics, one of which is that magnons can propagate with very low energy losses. This is because magnonic spin waves remove the need for electronic charge movement, thus removing the scattering of electrons and energy loss from heating issues associated with charge-based electronics. Another advantage of magnonics and the use of spin waves is that they have high tunability by adjusting magnetic and electric fields, spin currents, or thermal gradients \cite{ref2}. Magnons propagate by applying an electromagnetic field to a ferrimagnetic material, adjusting the magnetic field, manipulating the magnetic dipole moment, and changing the strength and direction of the spin wave \cite{ref3}. This is possible through voltage-controlled magnetic anisotropy (VCMA), which uses a DC or RF electric voltage field to manipulate the magnetic anisotropy \cite{ref4}. The Spin Seebeck effect shows that thermal gradients caused by ambient temperature or internal effects can be used to excite and manipulate spin waves \cite{ref5,ref6}. Furthermore, magnons have also been shown to be useful in memory information storage as magnon dark modes decouple from the RF cavity and have long-lifetime preservation. \cite{ref7}.

To amplify the unique advantages of magnonics, spin waves require thin films of 100 nm or less to propagate over long distances within small devices to mitigate propagation losses. Generally, thicker films have an increased defect density due to deposition or growth impurities, larger boundary grain formation, and defects in the crystal lattice, which results in more energy dissipation, reduced amplitude of the spin wave, and propagation losses due to scattering \cite{ref8}. Moreover, thin films of 100 nm or less thickness allow the manipulation of perpendicular magnetic shape anisotropy to realize magnon modes with out-of-plane magnetization, which would require very low bias fields \cite{ref9}. 

It is important to mention that thin film magnonic devices introduce their own set of challenges that require attention during device fabrications. First, they need to be highly crystalline, a process that usually involves a very high temperature during or after deposition or growth, which can lead to an interfacial reaction between the ferrimagnetic material and the host substrate. Crystallinity also affects the etching step in fabrication, as etching low-quality polycrystalline materials usually results in rough edges and poor aspect-ratio \cite{ref10}. This is significant because the etching of a highly single-crystalline material defines the geometry of the device, low side wall roughness, and high design control, all of which affect magnon propagation. Secondly, highly stoichiometric precursors are required to minimize impurities and scattering defects. Third, magnonic devices can operate only at low-power levels without creating excessive noise. The low-frequency noise for magnetic devices is dominated by random telegraph signal noise rather than 1/f noise for electronics \cite{ref11}. 

Thin film YIG has been shown to be the most promising material for nanoscale magnonic device applications \cite{ref12}, because it is compatible with nanofabrication as it can be deposited or grown to less than 100 nm \cite{ref13}, it is highly ferrimagnetic, and has the potential for magnon mode engineering due to its high spin density\cite{ref14}. The ferrimagnetism in YIG is driven by the crystal and chemical structure, which provides the material with many useful properties needed for magnonic devices. Ferrimagnets have two oppositely oriented magnetic moments with different magnitudes, resulting in a non-net zero magnetic moment \cite{ref15}. In YIG, magnons propagate through these non-zero magnetic moments inside the crystal lattice from the super exchange between the iron and oxygen ions \cite{ref16}. YIG has nearly no orbital angular momentum due to its magnetism because the \ce{Fe^{3+}} ions and spin are decoupled from the lattice, which contributes to the low damping of spin waves \cite{ref17}. However, vacancies in the \ce{Fe^{3+}} ions will result in reduced magnetizations, disrupting the super exchange interaction with the \ce{O^{2-}} ions, while oxygen vacancies will also hinder the lattice because they are the framework that holds the yttrium and iron ions together. This emphasizes the importance of having a uniform, stoichiometric, and preferably single-crystal YIG to reduce the propagation losses. Furthermore, YIG exhibits strong magneto-optical effects, where the polarization plane of the light rotates depending on the direction of the magnetic field \cite{ref18}.  

All these properties can be utilized in microwave and optical devices, which allow for the manipulation of electromagnetic waves at these frequencies. While crystalline YIG has typically only been realized on gadolinium gallium garnet (GGG) because of the lattice constant matching of YIG (a=12.367 Å) and GGG (a=12.383 Å) \cite{ref19},  GGG is not an ideal substrate because it is both difficult to process and can introduce additional magnetic damping due to the deleterious paramagnetic response\cite{ref20, ref21, ref22}. At low temperatures, the inner magnetic moments of the GGG can distort the magnetic properties required for magnons. This was caused by the geometric frustration of the GGG lattice, which did not properly align the magnetic moments\cite{ref23}. 

Given that the literature on achieving high-quality YIG thin films on other substrates is limited, we propose investigating  YIG on patterned Si/SiO$_2$ substrate \cite{ref24}. Nucleation seed patterning with YIG has been proven to aid in  achieving faster crystallization, which is useful for thermal budget \cite{ref25}. Recrystallization can be achieved using various high-temperature annealing techniques; here, we used an O$_2$ horizontal furnace \cite{ref26, ref27, ref28, ref29, ref30}. Since propagation losses can occur at the YIG to substrate interfaces, creating an interface layer of Si/SiO2 can minimize these effects and aid in the later suspension of the device. Suspended YIG devices have been shown to have low damping and linewidth in ferromagnetic resonance, proving that this fabrication method is compatible with CMOS devices \cite{ref31}. 
\section{Methodology}
\subsection{Material and Samples preparation}
\begin{figure}[htbp!]
\centering
\includegraphics[width=\linewidth]{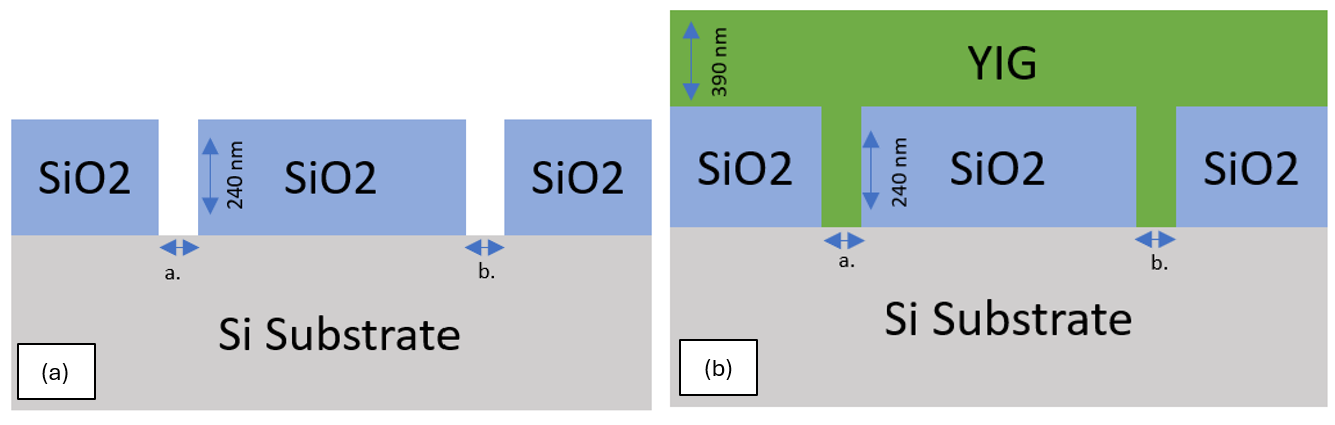}
\caption{ (a) Schematic of thermally grown and etched SiO$_2$. Dimensions of site a and b are 2 $\mu$m holes, separated either by 5 $\mu$m or 10 $\mu$m. (b) Schematic of process after YIG deposition.}
\label{fig:Figure 1}
\end{figure}

Si (100) wafer was used as the test substrate and subsequently cleaved into 1cm x 1cm samples. The samples were cleaned by rinsing with acetone, IPA, and DI water and then blown dry with N$_2$ gun to remove any particles during cleaving. The native oxide was then removed by bathing in 49\% diluted HF for 1 min each, followed by rinsing with DI water and blown drying with N$_2$ gun.
The samples were then inserted inside a horizontal furnace at 1100 °C under an O$_2$ atmosphere for 4 h to thermally grow 240 nm of SiO$_2$. Two sets of samples were fabricated and categorized into Batches 1 and 2. Batch 1 was a non-patterned substrate with SiO$_2$ and YIG deposited on top. Batch 2 is patterned SiO$_2$ with YIG deposited on the top.  The patterning was performed by maskless photolithography using a positive photoresist of AZ4330 and spin-coated at 6000 rpm. To print the patterned hole pairs, AZ4330 was exposed using a MLA 150 (Heidelberg Instruments) at a  dose of 280 $mJ/cm^2$ and developed for 1 min and 30 s. Patterned SiO$_2$ consisted of defined micro-hole pairs. These pairs consist of 2 $\mu$m diameters, separated by either 5 $\mu$m or 10 $\mu$m, and repeated every 50 $\mu$m. The SiO$_2$ was then etched using a Fluorine PlasmaTherm Takachi system and stopped at the Si/SiO$_2$ interface, as described schematically in Fig.~\ref{fig:Figure 1}a.

\begin{figure}[H]
\centering
\includegraphics[width=1\linewidth]{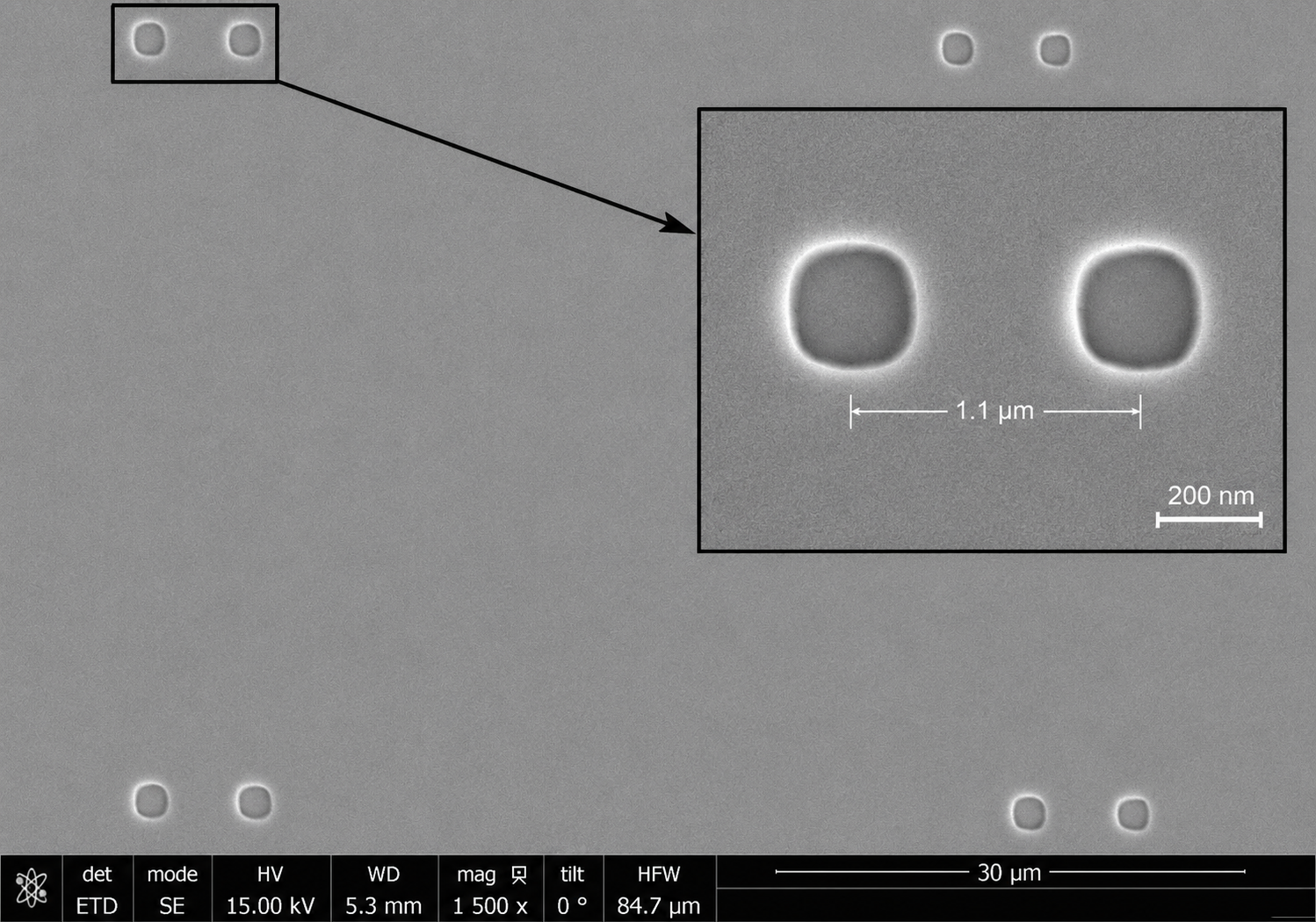}
\caption{ Top-down view of SEM image at 1500x of 2 $\mu$m x 5 $\mu$m sample at 4 location sites after YIG deposition. *Note* Contrast adjusted post imaging for contrast.}
\label{fig:Figure 2}
\end{figure}

The YIG target used for RF sputtering deposition has 99.99\% purity and dimensions of 76.2 mm in diameter and 3.175 mm in thickness, were purchased from Rearth Technology Company. YIG was deposited on the samples using a KJL PVD 75 sputter system. This was achieved at a vacuum base pressure of $1 \times 10^{-5}$ torr, 39.5 SCCM Argon gas flow, and 90 W RF power. Both batches 1 and 2 had a deposition time of 2 h or 7200 s, yielding 390 nm of YIG at a deposition rate of 0.054 nm/sec, which is schematically shown in Fig.~\ref{fig:Figure 1}b and in the top-down SEM image in Fig.~\ref{fig:Figure 2} for a 2 $\mu$m x 5 $\mu$m pattern. 
To study the optimal crystallization conditions, all the samples were annealed using a Lindberg S5359-BDS silicon diffusion horizontal furnace with O$_2$ gas at 6 SCCM flow. The furnace was set at three different temperatures, starting at 750 °C, 800°C, and 850 °C. At each temperature, the samples were annealed for 1 h, 2 h, and 3 h individually. Tab~\ref{tab:Table 1} lists these samples with their annealing parameters.
\begin{table}[htbp!]
    \centering
    \includegraphics[width=.85\linewidth]{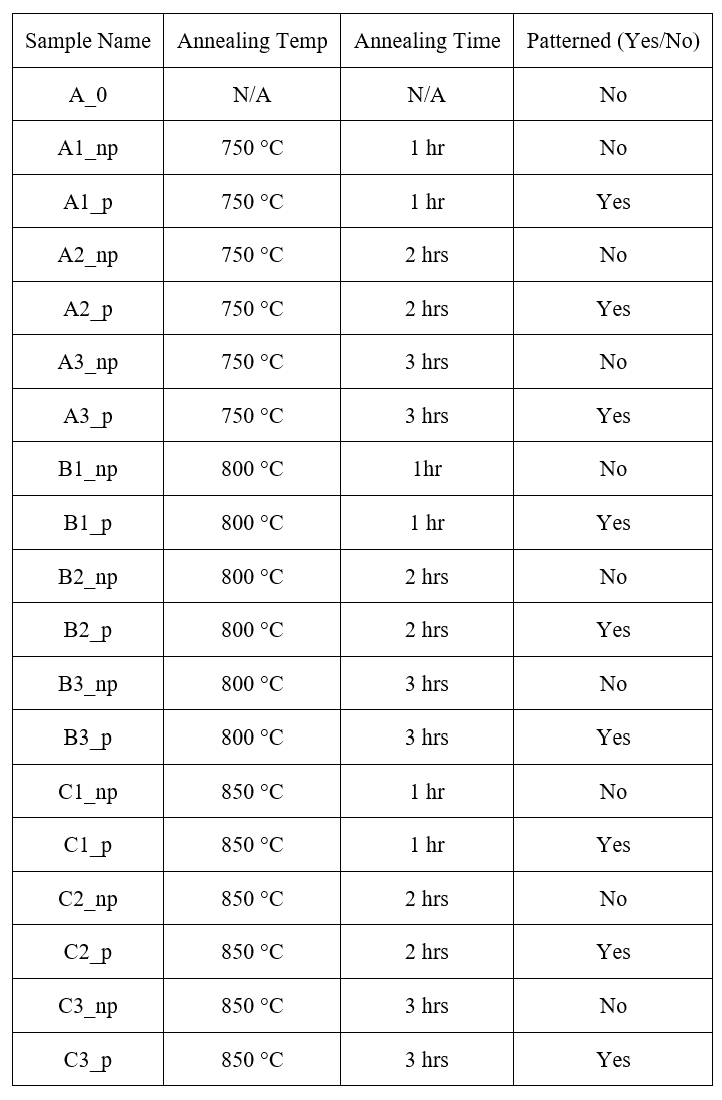}
    \caption{Sample names and annealing parameters.}
    \label{tab:Table 1}
\end{table}

\subsection{Defects and mitigation}
\begin{figure}[htbp!]
\centering
\includegraphics[width=\linewidth]{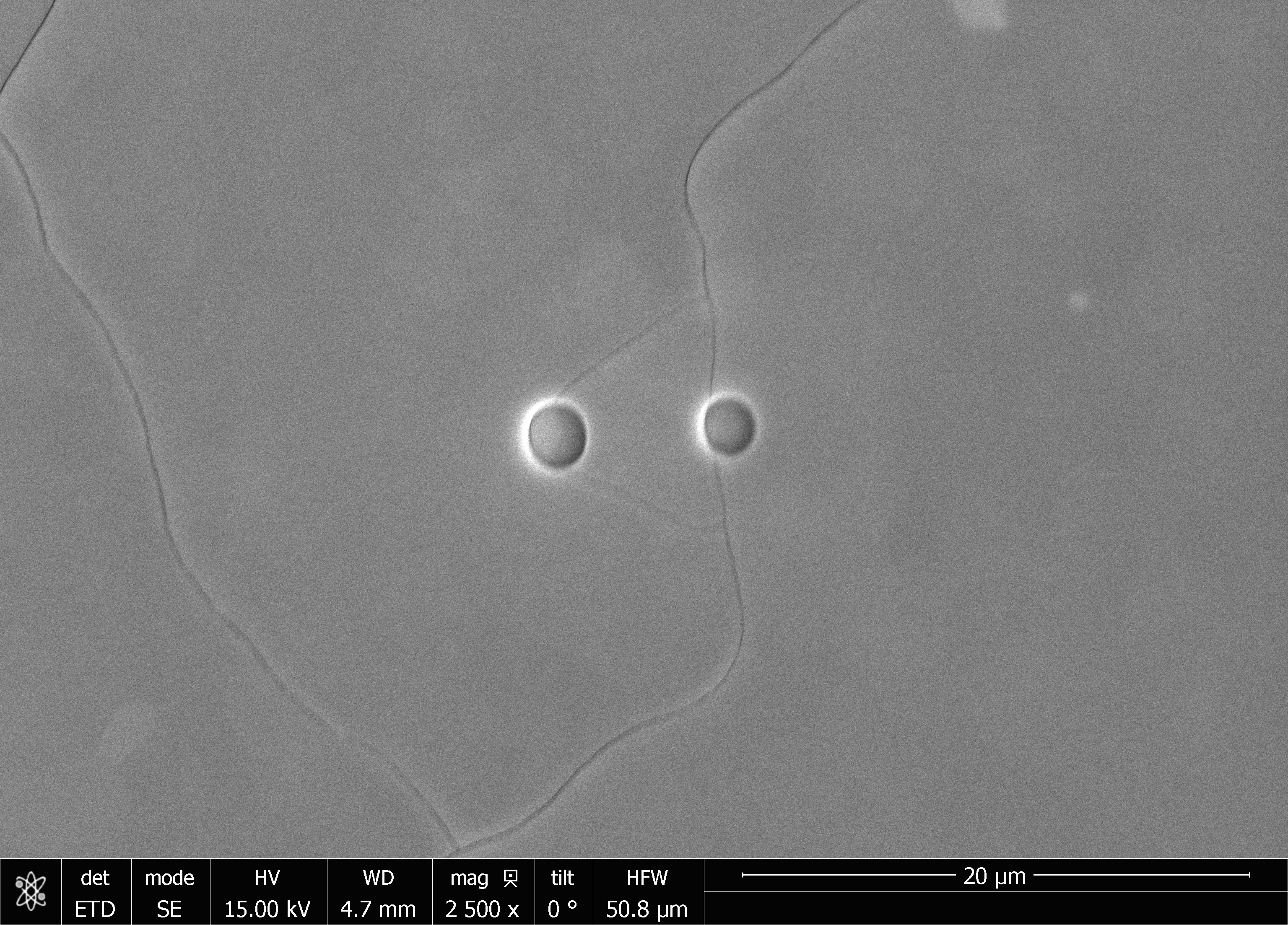}
\caption{Top-down SEM image (2500× magnification) of a 2$\mu$m × 5$\mu$m sample after YIG deposition and annealing at 800 °C for 1 h, showing crack formation.}
\label{fig:Figure 3}
\end{figure}

As shown in Fig.~\ref{fig:Figure 3}, one issue observed in the samples was cracking induced by tensile stress during annealing. Cracking is a well-documented challenge in YIG thin films and arises primarily from the mismatch in thermal expansion coefficients between YIG and Si/SiO$_2$ substrates, particularly for substrates other than GGG. Additionally, film thicknesses greater than approximately 100 nm further increase stress accumulation during high-temperature crystallization processes \cite{ref32}.

To mitigate this issue in future experiments, the use of a furnace with controlled increase and decrease rates is proposed to enable gradual heating and cooling during annealing. This approach is expected to reduce thermally induced stress and minimize crack formation. The furnace used in this study was selected to facilitate rapid annealing trials across a large sample set; however, it was not optimized to produce the highest-quality thin films.

The primary objective of this work was to investigate the effects of patterning on crystallization behavior and determine the optimal annealing temperature required to achieve high-quality crystallization in patterned YIG films. Consequently, a trade-off between processing throughput and film quality was necessary during this stage of experimentation.

\begin{figure}[htbp!]
\centering
\includegraphics[width=0.8\linewidth]{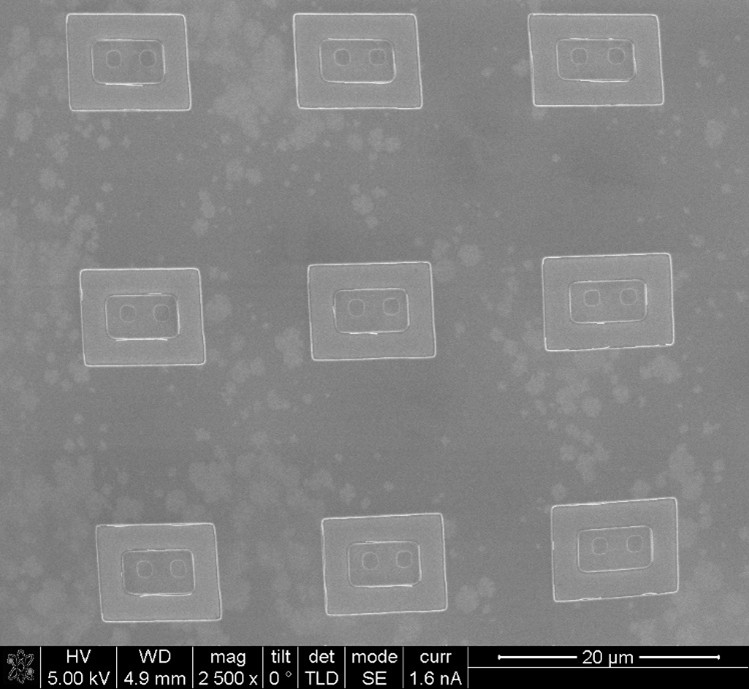}
\caption{Top-down SEM image (2500× magnification) of the updated isolated YIG array containing 1 µm × 2 µm hole-pair structures.}
\label{fig:Figure 4}
\end{figure}

An additional strategy to mitigate cracking involved modifying the pattern design. Previous studies have shown that isolated geometries on Si/SiO$_2$ substrates can significantly reduce or eliminate cracking \cite{ref33}. Based on this concept, an additional pattern layer was introduced around the border of the hole-pair structures, creating an array of isolated YIG regions separated by 12 µm in the y-direction and 9 µm in the x-direction, as shown in Fig.~\ref{fig:Figure 4}. Furthermore, the thickness of the YIG film was reduced to 100 nm, the thickness of SiO$_2$ was reduced to 50 nm, and the dimensions of the pair of holes were reduced to 1 $\mu$m × 2 $\mu$m while maintaining the same flow of the photolithography process.

\begin{figure}[htbp!]
\centering
\includegraphics[width=0.8\linewidth]{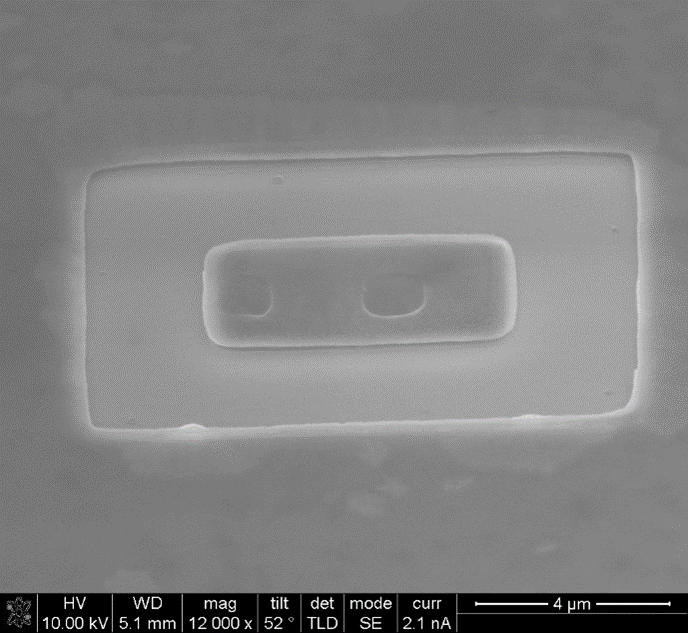}
\caption{Top-down SEM image (12000× magnification) of an isolated 1 µm × 2 µm YIG hole-pair structure after HF vapor suspension and annealing at 800 °C for 1 h, showing no evidence of cracking.}
\label{fig:Figure 5}
\end{figure}

This modified design introduces discontinuities within the YIG layer, thereby reducing stress propagation across the film. Following HF vapor suspension, the approach was found to eliminate cracking during annealing by creating separation within the SiO$_2$/YIG bilayer structure. Fig.~\ref{fig:Figure 5} shows a representative YIG hole-pair site after two minutes of HF vapor suspension followed by annealing at 800 °C for 1 h in the improved furnace, where no cracking is observed.
\

\begin{figure}[htbp!]
\centering
\includegraphics[width=0.8\linewidth]{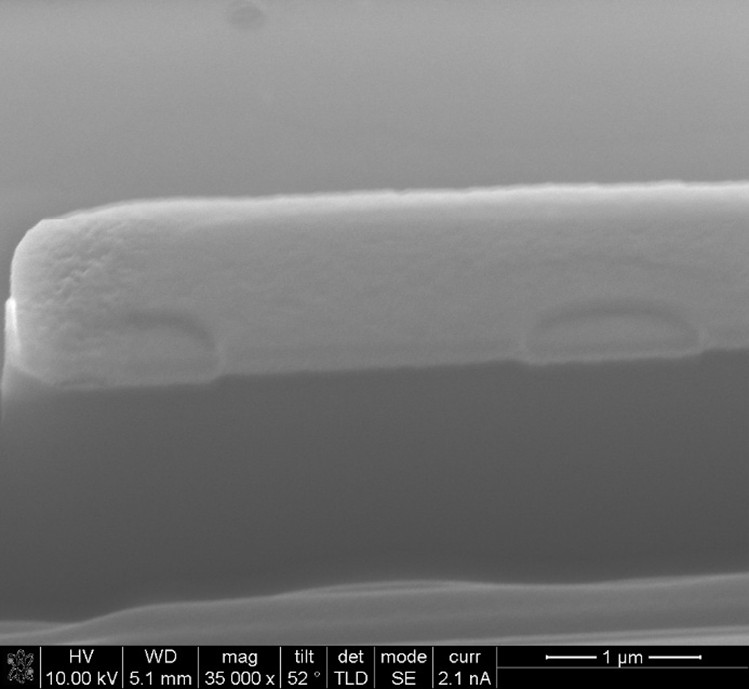}
\caption{SEM image of cross-sectional FIB (35000× magnification) of an isolated 1 µm × 2 µm YIG hole-pair structure after HF vapor suspension and annealing at 800 °C for 1 h, confirming approximately 50 nm of under-etching and the absence of cracking}
\label{fig:Figure 6}
\end{figure}

Fig.~\ref{fig:Figure 6} shows a cross-sectional FIB image of the same sample region presented in Fig. {5}. The image confirms that the HF vapor successfully etched approximately 50 nm beneath the hole-pair structure while maintaining structural integrity and preventing crack formation.

\begin{figure}[htbp!]
\centering
\includegraphics[width=0.8\linewidth]{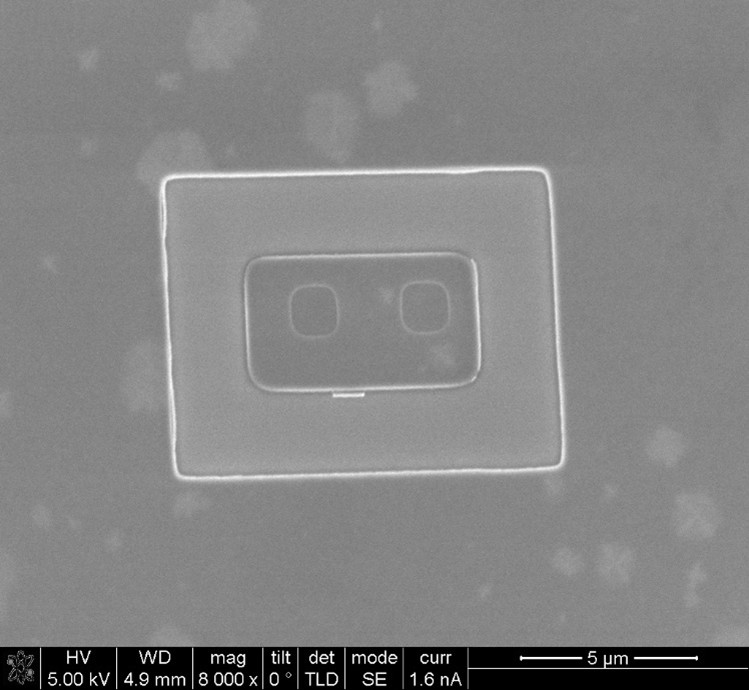}
\caption{Top-down SEM image (8000× magnification) of isolated 1 µm × 2 µm YIG hole-pair structures after annealing at 800 °C for 1 h without HF suspension, showing no observable cracking.}
\label{fig:Figure 7}
\end{figure}

Additionally, it was found that the isolated patterning design alone was sufficient to eliminate cracking without requiring HF vapor suspension. Annealing experiments performed on samples without HF suspension produced consistent results, indicating that crack suppression can be achieved primarily through the modified isolated-pattern geometry, as shown in Fig.~\ref{fig:Figure 7}. 

\section{Material Characterization}
To characterize the YIG material, several non-invasive techniques were used before and after crystallization. Energy Dispersive X-Ray Spectroscopy was performed using a FEI Nova Nanolab 600 SEM with an EDAX Apollo 40 detector and analyzed using the Genesis software.  EDS was used to determine the atomic composition using a ZAF correction algorithm. EDS measurements were done at 1000x magnification at 15 keV and 2 nA using area scans of 170 $\mu$m x 170 $\mu$m.

High-Resolution X-ray diffraction system from Rigaku with a Hypix detector was used to identify crystal phase. Parallel-Beam geometry was used using 2$\theta$-$\Omega$ area scan axis ranging from 20° to 90°, scan speed of 0.02 s, on continuous mode. Raw data were analyzed using PDXL 2.8, and the crystal phase and orientation were determined using the ICDD database. 

\begin{figure}[htbp!]
\centering
\includegraphics[width=\linewidth]{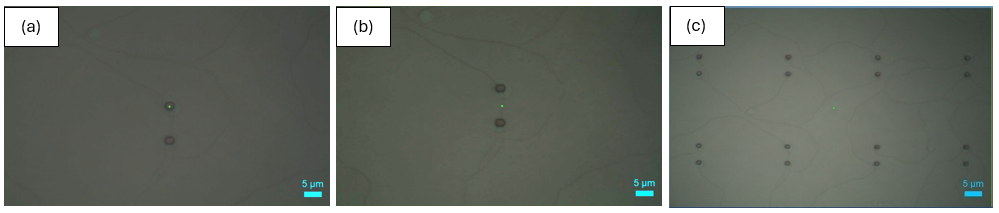}
\caption{(a) Raman microscope and laser measurement spot inside hole. (b) Between hole pairs. (c) 25 $\mu$m away from hole pairs. *Note* Contrast and brightness adjusted post imaging.}
\label{fig:Figure 8}
\end{figure}

Raman measurements were performed to identify vibrational modes and structural analysis using a LabRAM HR Evolution Raman Spectrometer (Jobin Yvon Technology). The data were then analyzed using LabSpec6 software. Measurements were taken at multiple averaged points throughout the sample from 150 $cm^{-1}$ to 750 $cm^{-1}$ range, at 1800 gr/mm, 442 nm wavelength laser, 100x magnification, with 100\% laser intensity, and 30 s scan times. Achieving a spectral resolution of about 0.65 $cm^{-1}$, for both the non-patterned and patterned samples. For the non-patterned samples, these measurements were taken randomly inside the 1 cm x 1 cm samples. For the patterned samples, the area measurements were taken inside a hole, between a hole pair, and 25$\mu$m away from the hole pair, as shown in Fig \ref{fig:Figure 8}.

Spectroscopic Ellipsometry (SE) measurements were conducted using a J.A Woollam M-2000 to determine both the thickness and optical constants. To characterize thin-film YIG,  a modified B-spline model was created using the initial parameters derived from the stoichiometric YIG target. Additionally, SE has been employed to extract high-frequency dielectric constants as a function of wavelength, ranging from 200 to 1700 nm. 
\section{Results and discussion}
\subsection {EDS}
Electron beam spot size measurements were 170 $\mu$m x 170 $\mu$m, and an accelerating voltage of 15 keV was used. Three measurements were taken and averaged for the sample deposited at room temperature and for non-patterned and patterned samples annealed at 750 °C, 800 °C, and 850 °C. Results shown in Tab.~\ref{tab:Table 2} - Tab.~\ref{tab:Table 4}  for the atomic percentage and atomic weight value suggest that as the temperature and annealing time increase, the YIG seems to vary slightly in stoichiometry as compared to the as-deposited sample, likely due to varied RF sputtering conditions during individual depositions. 
The trend shows that the values are still somewhat maintained, although we have slightly more iron than ideal, and the stoichiometry remains constant throughout the annealing temperatures and times. Therefore, the stoichiometric amorphous and crystalline YIG were determined to be the same throughout annealing. Since there was no significant change in the oxygen atomic percentage, it was also determined that there is no major interaction between the YIG and Si/SiO$_2$ substrate, and it can be said that there was no absorption or desorption of oxygen from the environment. The EDS results were verified using other YIG samples of varying thicknesses to confirm that the software quantitative analysis of the atomic percentages was consistent. 
The observed off-stoichiometry is attributed primarily to the RF sputtering conditions. In this study, deposition was performed using only Ar as the sputtering gas. However, incorporating an Ar:O$_2$ gas mixture would increase the oxygen partial pressure during growth, potentially improving film stoichiometry and reducing oxygen deficiency. Previous studies have reported improved compositional control using Ar:O$_2$ ratios as high as 80:20 \cite{ref13}, although a range of gas mixtures has been explored depending on system configuration and target composition.
\begin{table}[htbp!]
    \centering
    \includegraphics[width=1\linewidth]{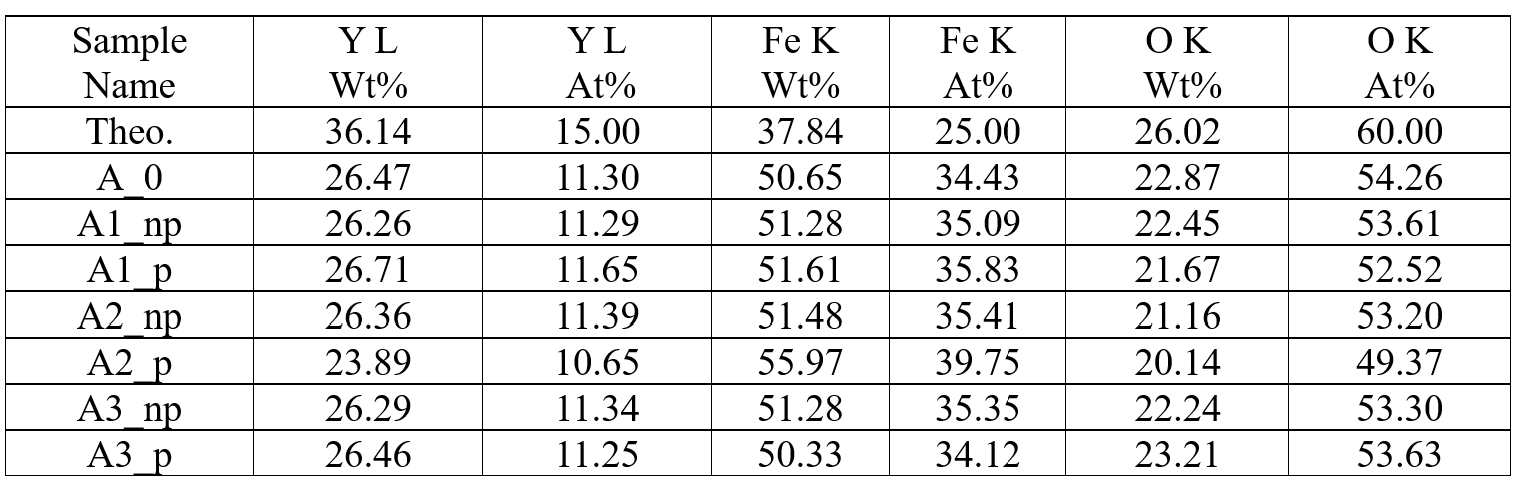}
    \caption{EDS Weight and Atomic Composition \% results for 750°C annealed samples.}
    \label{tab:Table 2}
\end{table}
\begin{table}[htbp!]
    \centering
    \includegraphics[width=1\linewidth]{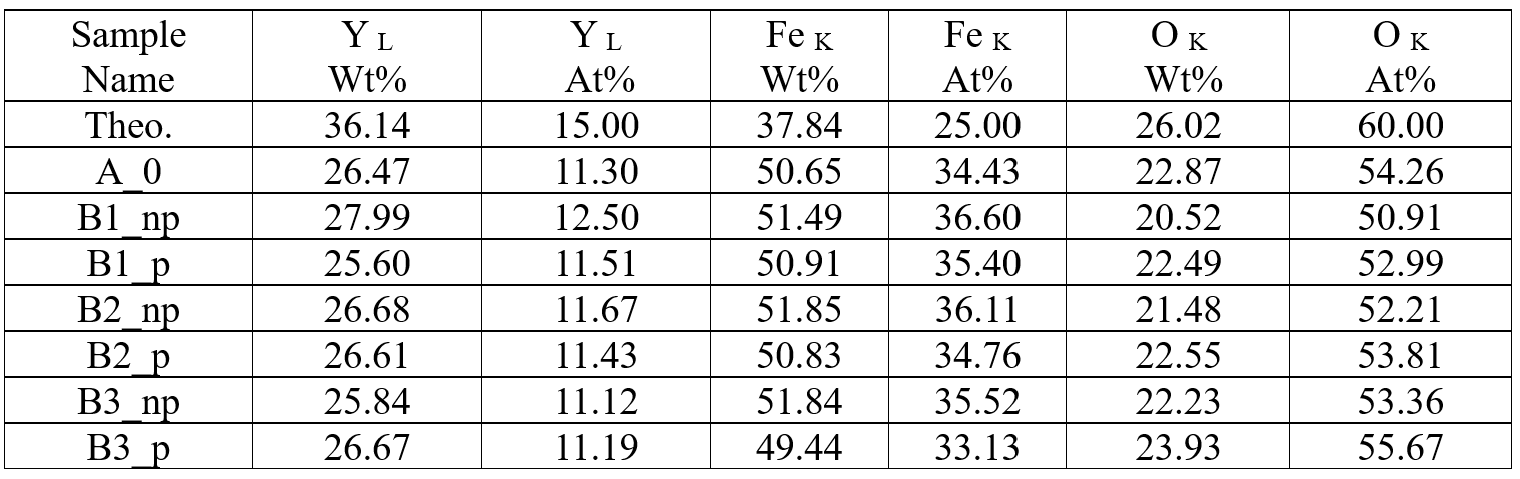}
    \caption{EDS Weight and Atomic Composition \% results for 800°C annealed samples.}
    \label{tab:Table 3}
\end{table}
\begin{table}[htbp!]
    \centering
    \includegraphics[width=1\linewidth]{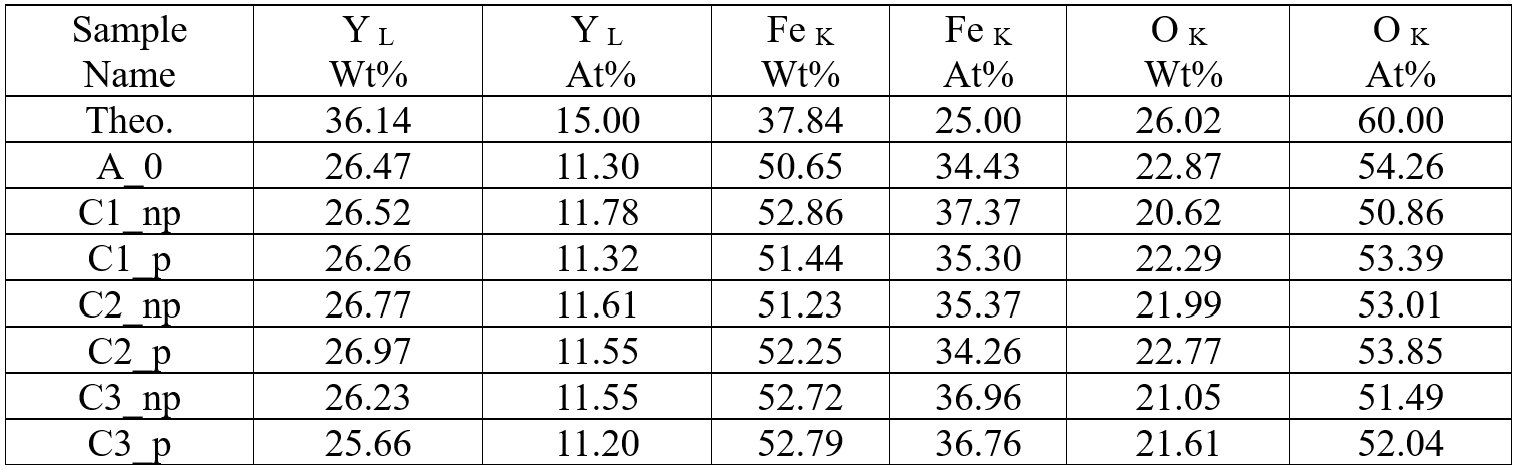}
    \caption{EDS Weight and Atomic Composition \% results for 850°C annealed samples.}
    \label{tab:Table 4}
\end{table}

The observed off-stoichiometry is attributed primarily to the RF sputtering conditions. In this study, deposition was performed using only Ar as the sputtering gas. However, incorporating an Ar:O$_2$ gas mixture would increase the oxygen partial pressure during growth, potentially improving film stoichiometry and reducing oxygen deficiency. Previous studies have reported improved compositional control using Ar:O$_2$ ratios as high as 80:20 [13], although a range of gas mixtures has been explored depending on system configuration and target composition.
\subsection{XRD}
\begin{figure*}
\includegraphics{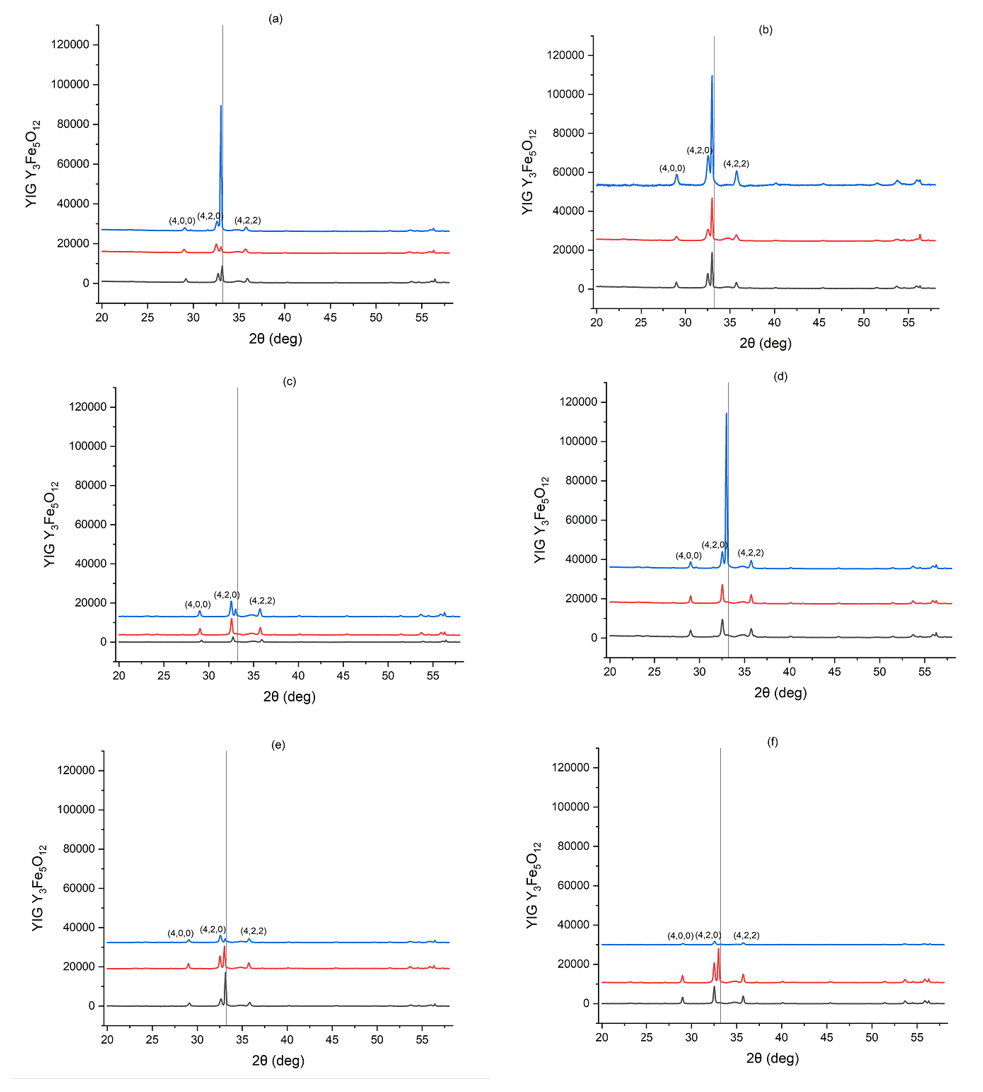}
\caption{\label{fig:wide}(a) XRD patterns for 750 °C anneals for non-patterned YIG, (b) XRD patterns for 750 °C anneals for patterned YIG, (c) XRD patterns for 800 °C anneals for non-patterned YIG, (d) XRD patterns for 800 °C anneals for patterned YIG, (e) XRD patterns for 850 °C anneals for non-patterned YIG, (f) XRD patterns for 850 °C anneals for patterned YIG. XRD patterns are colored as follows: black patterns indicate 1 h annealing, red patterns indicate 2 h annealing, and blue pattern indicates 3 h annealing.}
\label{fig:Figure 9}
\end{figure*}

The XRD patterns in Fig.~\ref{fig:Figure 9}(a-f), along with the corresponding position, integrated intensity, and FWHM of 2$\theta$ in Tab.~\ref{tab:Table 5} were analyzed using a fit with PDXL software and the ICDD database \cite{ref34}. The analysis confirmed the formation of YIG peaks at 2$\theta$ $\approx$ 29.0°, 32.6°, and 35.7°, corresponding to the cubic orientations (400), (420), and (422), respectively. These peaks were observed across all samples, with only minor shifts of $\le$ 0.2°, suggesting slight lattice distortions as a function of the annealing conditions.

The intensity and broadening of the diffraction peaks differ significantly between the non-patterned (\_np) and patterned (\_p) samples. Patterned samples generally exhibit sharper FWHM and more intense peaks, indicating improved crystallinity and a larger crystallite size. In contrast, the broader and lower-intensity peaks of non-patterned samples suggest a smaller grain size and possible lattice strain. To identify the optimal annealing conditions, both the peak intensity and broadening (FWHM) were compared.

For patterned samples, as shown in Fig.~\ref{fig:Figure 9}(b), annealed at 750 °C, the YIG (420) peak intensity increases with annealing time: 1420 (cps °) (A1\_p, 1 h annealing), 1671 cps° (A2\_p, 2 h annealing), and 3853 cps° (A3\_p, 3 h annealing). However, these samples also show relatively broad FWHM values (0.22–0.35°), indicating that the single-crystal phase is still incomplete. At 800 °C, shown in Fig.~\ref{fig:Figure 9}(d) the maximum peak intensity of 1788 cps° is achieved after 1 h, but the intensity decreases at longer times, with the corresponding FWHM of 0.22°. By comparison, out of all the non-patterned samples, annealing at 800 °C for 3 h, as shown in Fig.~\ref{fig:Figure 9}(c), reached its highest intensity of 1652 cps° with a much narrower FWHM of 0.13°.

Since there is a trade-off between crystallinity and the thermal budget required for annealing (because YIG is prone to cracking under heat stress), we conclude that patterned samples reach optimal crystallization more quickly than non-patterned ones, with the best combination of peak intensity and FWHM obtained for patterned samples annealed 800 °C annealing for 1 h. These conditions reduce the required thermal budget, as increasing the temperature by just 50 °C,  from 750 °C to 800 °C eliminates the need for an additional three or more hours of annealing required for 750 °C. Otherwise, a two-step annealing process (e.g., 3 h at 750 °C followed by 3 h at 500 °C) would be necessary to further avoid cracking \cite{ref35}. Furthermore, increasing the temperature further to 850 °C, seen in Fig.~\ref{fig:Figure 9}(e) and (f), does not improve crystallinity; instead, it leads to thermal damage which results in more cracking of YIG, as reflected by the significantly reduced intensities (383–1747 cps°) for patterned samples.

\begin{figure}[htbp!]
\centering
\includegraphics[width=1\linewidth]{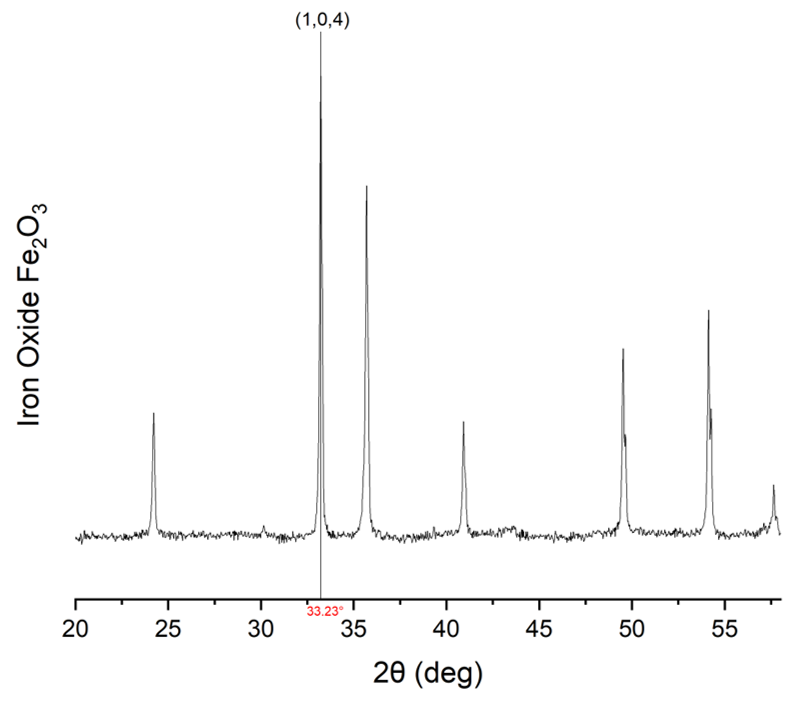}
\caption{XRD data for hematite with crystal peak (1,0,4) labeled located at 33.23° [36].}
\label{fig:Figure 10}
\end{figure}

In addition to YIG, a secondary peak near 2$\theta$ $\approx$ 33.0° was observed for some samples. This peak corresponds to the (104) plane of rhombohedral \ce{Fe_2O_3} (hematite), indicating the incomplete phase formation of YIG or unreacted iron oxide. Non-stoichiometric YIG, particularly when annealed at low temperatures (<1400 °C) in an oxygen-rich environment, can result in segregated phases of \ce{YFeO_3} (YIP) and \ce{Fe_2O_3} (hematite) \cite{ref36,ref37}. Given that our initial deposition was off-stoichiometric and rich in iron, we determined this peak to be \ce{Fe_2O_3}. PDXL also consistently identified this peak as hematite (104), and this peak matched the XRD reference data for \ce{Fe_2O_3} \cite{ref38}. This can be observed in  Fig.~\ref{fig:Figure 10}, where the \ce{Fe_2O_3} (104) peak is located at $\sim$33.23° and is marked by a line. The same identification is listed in the XRD data Fig.~\ref{fig:Figure 9}(a–f) around $\sim$ 33.0° and is tabulated in Tab.~\ref{tab:Table 5} under the \ce{Fe_2O_3} (104) rows.
\clearpage
\begin{table}[htbp!]
    \centering
    \includegraphics[width=1\linewidth]{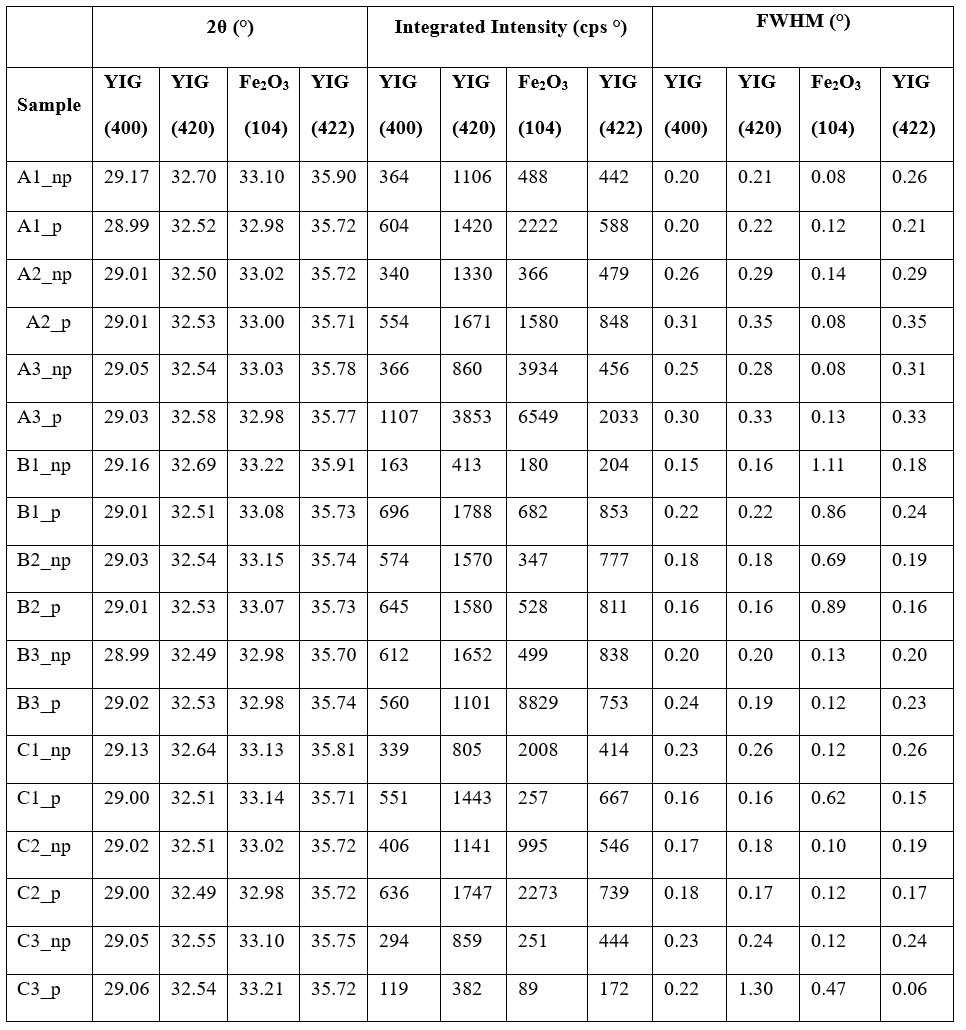}
    \caption{XRD 2$\theta$ position (°), integrated intensity (cps°), and FWHM (°) data for the YIG (400), (420), \ce{Fe_2O_3} (104) and YIG (444) for all non-patterned and patterned samples. Table is organized by increasing 2$\theta$. }
    \label{tab:Table 5}
\end{table}
\subsection {Raman}
Three Raman peaks were identified using fit analysis with Origin 2025 software and were assigned to the vibrational \ce{T_{2g}} mode of \ce{Y^3+} and \ce{[FeO_4]^5-} in the dodecahedral lattice, the \ce{E_g} mode of \ce{Y^3+} and \ce{[FeO_4]^5-} in the dodecahedral lattice, and the \ce{E_g} internal mode of the free \ce{FeO_4} tetrahedra \cite{ref26,ref27,ref28,ref39,ref40,ref41} located at approximately 222 $cm^{-1}$, 287 $cm^{-1}$, and 405 $cm^{-1}$, respectively. These assignments were made by comparison with the literature values listed in Tab.~\ref{tab:Table 6} where the data are organized by bulk or thin film, growth method, and vibrational mode positions $cm^{-1}$. Our peaks are slightly shifted in frequency compared to the referenced data, which is due to two reasons: one, we have excess iron oxide as demonstrated by EDS and XRD, and second, we have less yttrium in the YIG than expected in stoichiometry. This change in yttrium causes a shift in frequency, as previously demonstrated (Brasil et al, 2025)\cite{ref42}.

All samples were measured and analyzed; however, when comparing the results to the XRD analysis, we specifically focused on the Raman data for the samples annealed at 800 °C, since XRD confirmed this to be the optimal annealing temperature. Fig.~\ref{fig:Figure 6} shows the Raman spectra of the non-patterned samples annealed at 800 °C for 1, 2, and 3 h. Fig.~\ref{fig:Figure 7} presents the Raman spectra at specific measurement locations for the patterned sample annealed at 800 °C for 1 h: 25 $\mu$m away from a hole pair, between the hole pair, and inside the hole pair, as shown by the green dot in Fig.~\ref{fig:Figure 3}. 

For the \ce{T_{2g}} mode in the non-patterned 800 °C annealed samples (Fig.~\ref{fig:Figure 6}), the overall intensity increased with annealing time, reaching a maximum of 414.85 cps at 3 hours. The FWHM progressively narrowed, reaching a minimum of 9.22 $cm^{-1}$, consistent with the expectations for non-patterned YIG. However, XRD revealed that such extended annealing is not ideal due to heat-induced stress. 

\begin{figure}[H]
\centering
\includegraphics[width=.9\linewidth]{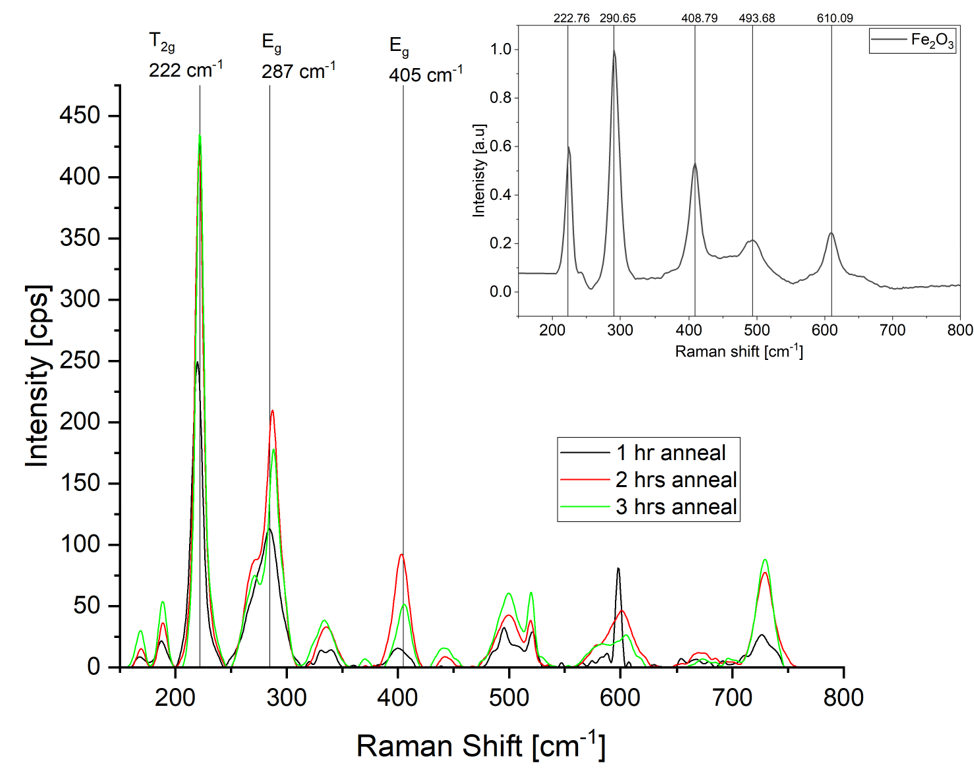}
\caption{: Raman peaks for 1 h, 2 h, and 3 h at 800° C anneals for non-patterned YIG.}
\label{fig:Figure 11}
\end{figure}

The inset in Fig.~\ref{fig:Figure 11} shows the \ce{Fe_2O_3} Raman data \cite{ref43}, revealing similar peak positions around our identified YIG peaks at $\approx$ 222 $cm^{-1}$, $\approx$ 290 $cm^{-1}$, and $\approx$ 410 $cm^{-1}$. This is mainly because to YIG and Hematite have the same vibrational building blocks, and Raman spectra dominated by the Fe-O bonds are present in both materials. The key differences are the intensity of each peak, where YIG has more Raman-active modes due to its complex structure, and hematite has fewer and sharper peaks. 

\begin{figure}[H]
\centering
\includegraphics[width=0.8\linewidth]{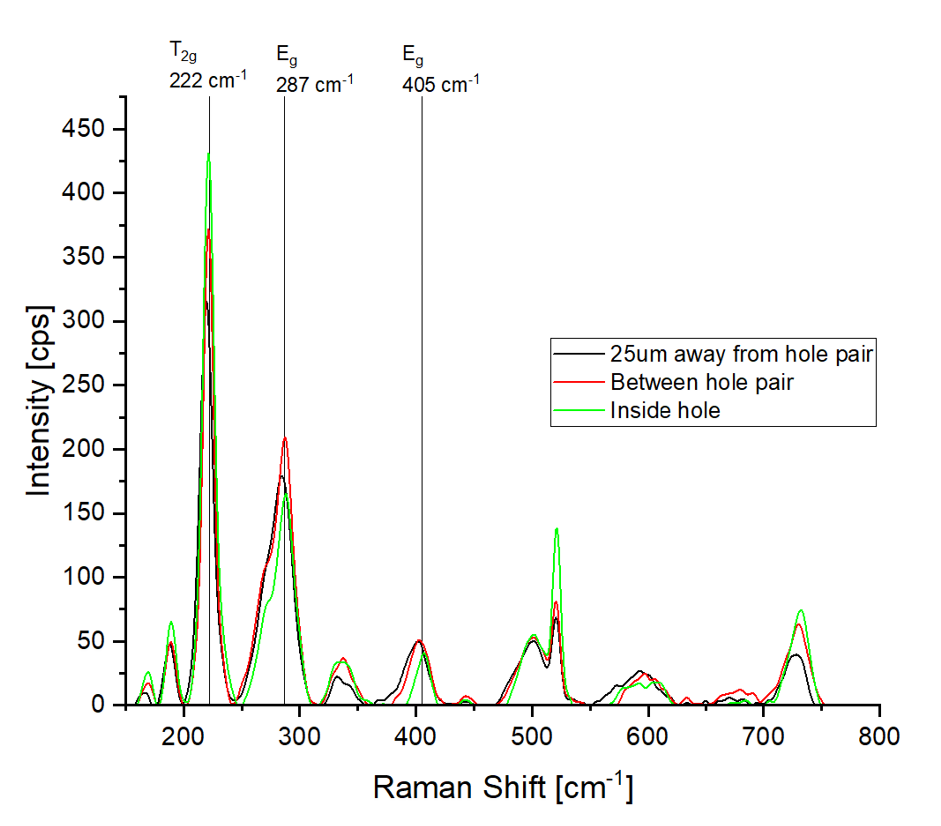}
\caption{One hour 800° C anneal for patterned YIG, 25 $\mu$m away, between hole, and inside hole.}
\label{fig:Figure 12}
\end{figure}

Examining the same \ce{T_{2g}} mode in patterned YIG (Fig.~\ref{fig:Figure 12}) for the 1 hour 800 °C anneal, the highest intensity was observed inside the hole, with a value of 414.66 cps and an FWHM of 12.50 $cm^{-1}$. The intensity decreased progressively with distance from the hole pair, reaching a minimum of 301.85 cps and a FWHM of 13.27 $cm^{-1}$ at 25 $\mu$m away. This indicates that crystallization propagates outward from the patterned holes, likely due to seed nucleation. 

Based on the tabulated Raman data for Si \cite{ref44} and \ce{SiO_2} \cite{ref45}, we confirm that our identified YIG Raman peaks are not associated with the substrate; however, the peak seen at approximately 522 $cm^{-1}$ originates from the Si substrate and is seen with the highest intensity on patterned samples, in Fig.~\ref{fig:Figure 11} inside the hole, due to its closest proximity to the substrate. Changes in the Raman spectra from sample to sample were also not related to thickness, as we consistently deposited the same thickness of approximately 390 nm thin film YIG. 

As expected, these Raman results agree with the XRD results, confirming that patterned samples fully crystallize at 800 °C after only 1 h, in contrast to the non-patterned samples. Moreover, the data demonstrate that crystallization begins inside the patterned holes and propagates outward.
\begin{table}[H]
    \centering
    \includegraphics[width=1\linewidth]{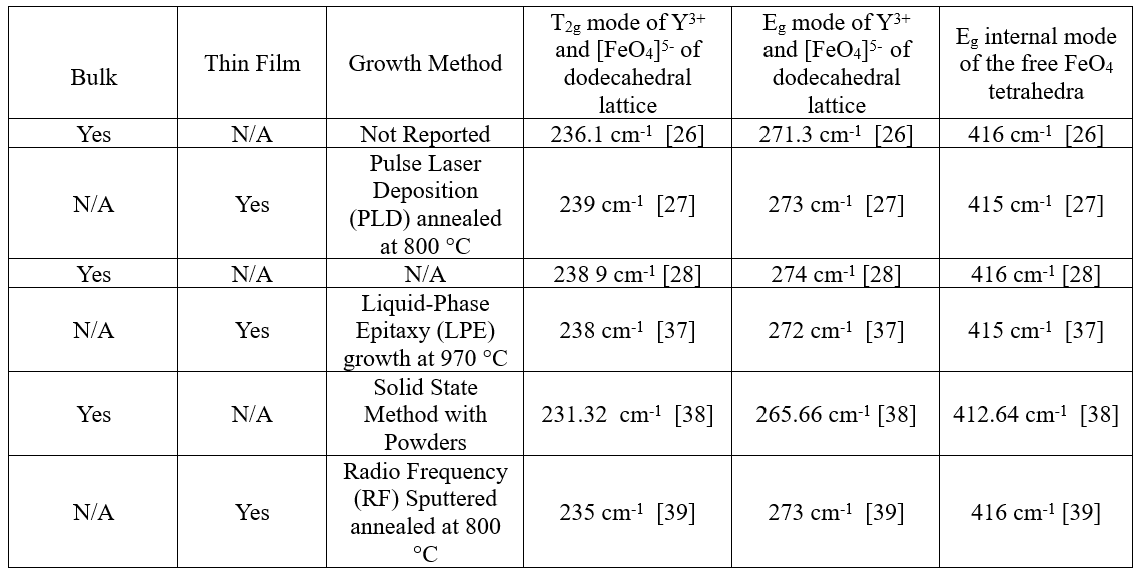}
    \caption{YIG Raman Peak modes from literature.}
    \label{tab:Table 6}
\end{table}
\begin{table}[H]
    \centering
    \includegraphics[width=1\linewidth]{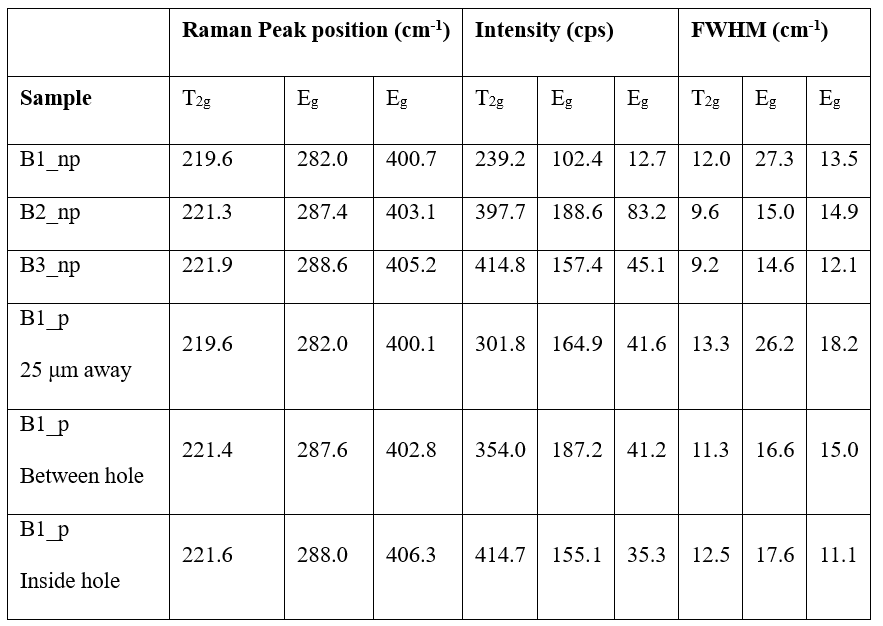}
    \caption{: Raman peak position ($cm^{-1}$), intenisty (cps), and FWHM ($cm^{-1}$) for three identified peaks of vibrational modes \ce{T_2g} mode of \ce{Y^3+} and \ce{[FeO_4]^5-} of dodecahedral lattice, \ce{E_g} mode of \ce{Y^3+} and \ce{[FeO_4]^5-}  of dodecahedral lattice, and \ce{E_g} internal mode of the free \ce{FeO_4} tetrahedra, for non-patterned YIG for 1 hour, 2 hours, and 3 hours annealing at 800 °C and patterned YIG at 800 °C for 1 hour annealing for measurements locations of  25 $\mu$m away from hole pair, between hole pair, and inside hole. }
    \label{tab:Table 6}
\end{table}
\subsection{Ellipsometry}
Ellipsometry measurements were used to confirm the YIG thin film thickness, which was approximately 390 nm for the 7200 s RF sputtering deposition. Refractive index (n) values were obtained for all samples over a wavelength range of 200–1700 nm. However, the analysis focused on the 800 °C non-patterned and patterned samples annealed for 1 h, 2 h, and 3 h.

\begin{figure}[H]
\centering
\includegraphics[width=\linewidth]{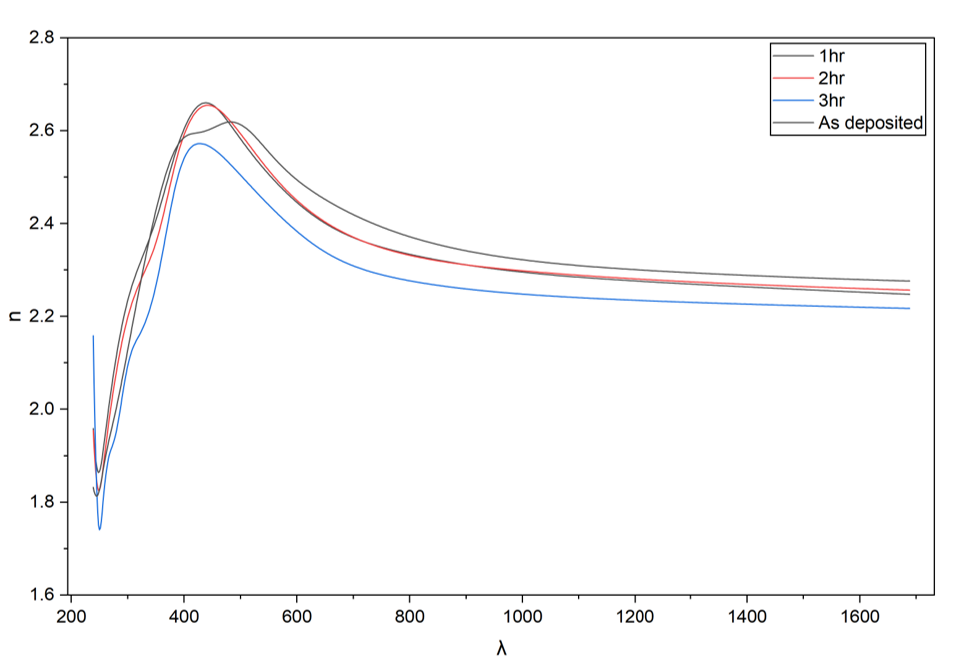}
\caption{Index of refraction n vs wavelength data for 800 °C anneals on non-patterned samples.}
\label{fig:Figure 13}
\end{figure}

As shown in Fig.~\ref{fig:Figure 13}, the as-deposited (non-annealed) YIG film exhibits a non-oscillatory n curve with a max peak at 485 nm and n$\approx$2.6. This behavior is characteristic of amorphous or low-density YIG. Upon annealing, the n value of the non-patterned YIG increased, indicating film crystallization. The 1 h annealed non-patterned sample shows the highest refractive index of n$\approx$2.66 at 437 nm. Prolonged annealing for 2 or 3 h results in a reduction of n, which is consistent with the XRD data and suggests heat-induced stress, as previously discussed.

\begin{figure}[H]
\centering
\includegraphics[width=1.1\linewidth]{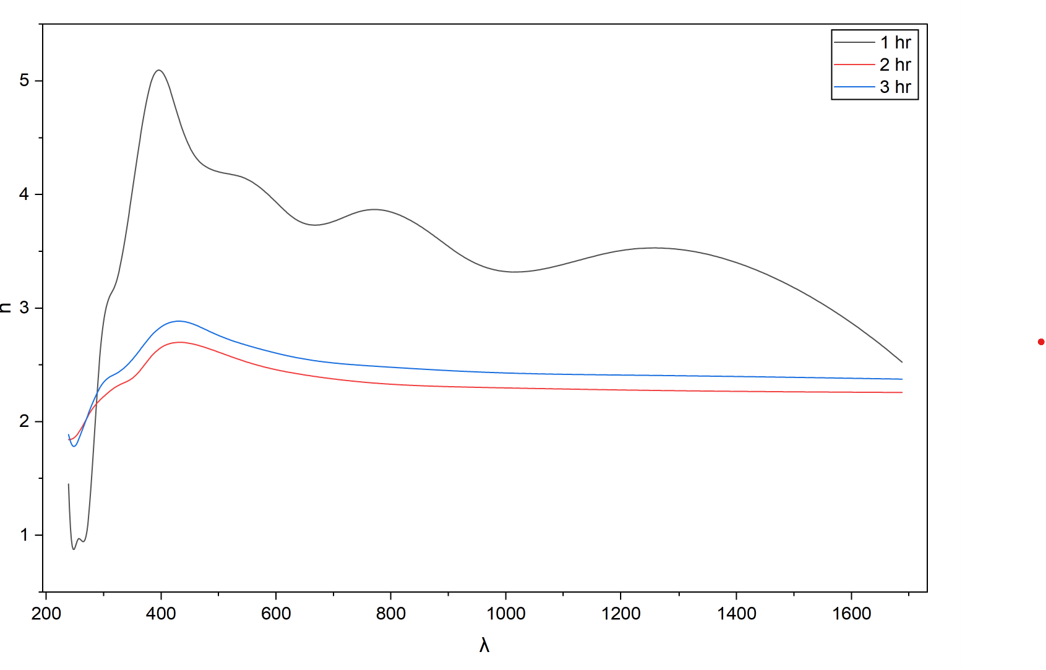}
\caption{Index of refraction n vs wavelength data for 800 °C anneals on patterned samples.}
\label{fig:Figure 14}
\end{figure}

For the patterned YIG samples shown in Fig.~\ref{fig:Figure 14}, the 1 h sample exhibits unusually high and oscillatory n values, with a peak at 395 nm of n$\approx$ 5.1. This behavior may result from the optical interference caused by the coexistence of two segregated phases, YIP and YIG, as identified in the XRD analysis, as well as contribution from the substrate as cracking is introduced during annealing. The n curves annealed for 2 h and 3 h exhibited lower refractive index values of n$\approx$ 2.7 and $\approx$ 2.88, respectively. However, these values are still higher than those of the non-patterned samples and are consistent with the XRD and Raman results, proving that the pattern samples crystallized faster.  
\section{Conclusion}
In summary, we demonstrated the crystallization of RF sputtered YIG thin films on Si/SiO$_2$ using a combination of non-invasive characterization methods and examined their evolution from amorphous to crystalline phases. We analyze two types of structures, one pure thin film of YIG on Si/SiO$_2$, and the other on patterned Si/SiO$_2$, to create crystallization seeds directly on Si.

Our results indicate that the patterned sample annealed at 800 °C for 1 h promoted faster crystallization. This is a greater advantage than that previously reported in the literature for two reasons. First, we can reduce the thermal budget required to crystallize YIG, and second, by removing SiO$_2$, we can create a suspended YIG crystalline YIG on Si. This methodology allows compatible Si magnon devices to be fabricated using a relatively low-temperature process. 

Annealing of continuous YIG thin films and single-layer patterned structures, including hole-pair geometries, was found to induce cracking due to thermally generated stress. The introduction of an additional patterning layer to form isolated YIG structures effectively eliminated crack formation during annealing. These structures may be annealed directly or processed with HF vapor to partially remove the underlying SiO$_2$ sacrificial layer before annealing. Although the fabrication approach successfully mitigated cracking, further optimization of the sputtering process is required to achieve stoichiometric YIG. Following resolution of both the cracking and stoichiometry challenges, magnetic characterization will be conducted to assess spin-wave behavior and device performance, as structural defects and compositional nonuniformities can significantly impact magnetic properties and spin-wave propagation.

\section{Supplementary documentation}
\begin{figure}[H]
\centering
\includegraphics[width=\linewidth]{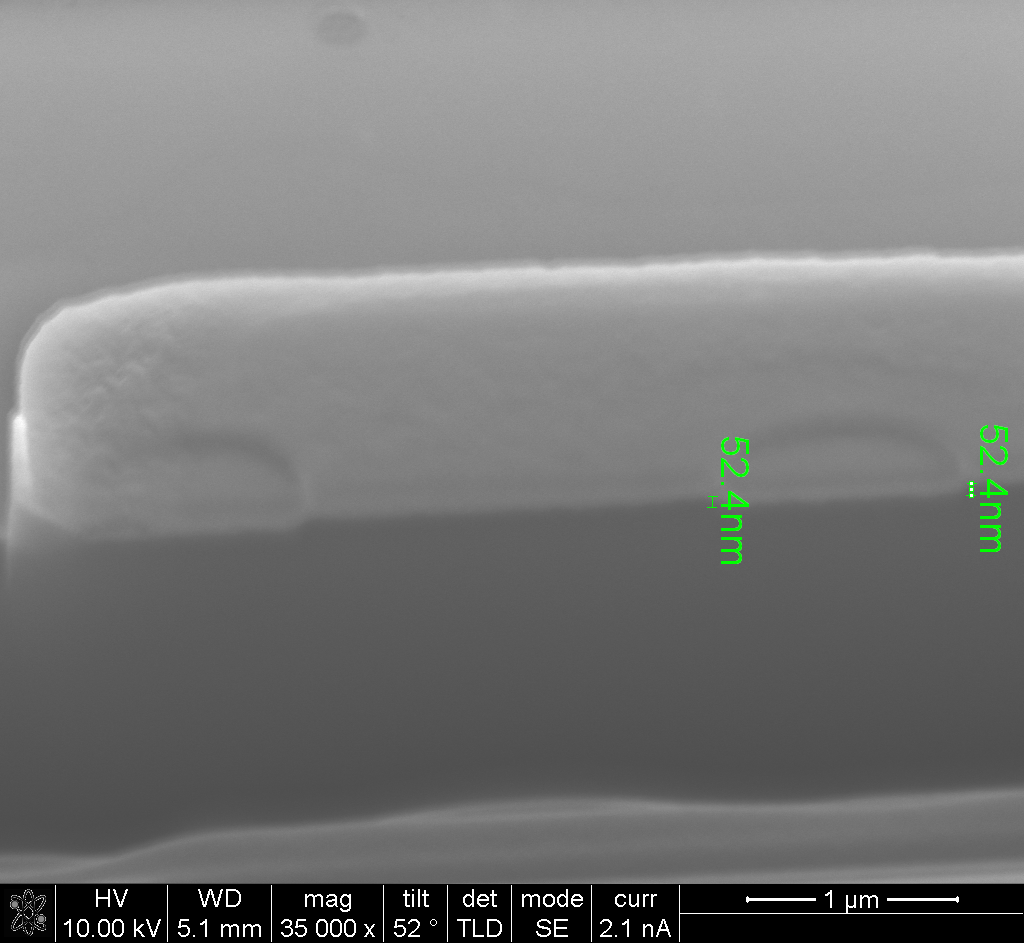}
\caption{ SEM image of Cross-sectional FIB (35000× magnification) of an isolated 1 $\mu$m × 2 $\mu$m YIG hole-pair structure after HF vapor suspension and annealing 800 °C for 1 h, confirming approximately 50 nm of under-etching and the absence of cracking with measurements.}
\label{fig:Figure 15}
\end{figure}

Fig.~\ref{fig:Figure 15} reproduces Fig.~\ref{fig:Figure 6} from the main text with additional measurements indicating that approximately 50 nm of SiO$_2$ was etched during HF vapor treatment. The added measurement markers cover the etched region; therefore, the figure is included here to provide further clarification.

\section{Acknowledgments}
This work was performed, in part, at the Center for High Technology Materials (CHTM) and Center for Integrated Nanotechnologies (CINT), an Office of Science User Facility operated for the U.S. Department of Energy (DOE), and Office of Science Sandia National Laboratories. Sandia National Laboratories is a multimission laboratory managed and operated by National Technology \& Engineering Solutions of Sandia, LLC, a wholly owned subsidiary of Honeywell International Inc., for the U.S. Department of Energy’s National Nuclear Security Administration under Contract DE-NA0003525. This paper describes objective technical results and analysis. Any subjective views or opinions that might be expressed in the paper do not necessarily represent the views of the U.S. Department of Energy or the United States Government.

We acknowledge Isaac Strictlin and Caleb Annan from the Busani research group for their support in GDS file creation and ellipsometry data collection and analysis. We also acknowledge the use of the Horiba LabRam Microscope in the Cavallo Research Laboratory at CHTM/UNM. Specifically, we thank Dr. Divya Prakash for their expert training on the tool and assistance with initial data acquisition. In addition, we are grateful to Dr. Francesca Cavallo for insightful discussions regarding the acquired Raman spectra.
\nocite{*}
\bibliography{references}
\end{document}